\documentclass[aps,twocolumn]{revtex4}
\usepackage{amsmath,amssymb}

\usepackage{epsfig}
\usepackage{graphicx}

\newcommand{\be}{\begin{equation}}
\newcommand{\ee}{\end{equation}}
\newcommand{\bea}{\begin{eqnarray}}
\newcommand{\eea}{\end{eqnarray}}
\newcommand{\nn}{\nonumber}

\begin{document}

\title{Gravitating defects of codimension-two}

\author{Georgios Kofinas\footnote{gkofin@phys.uoa.gr} and
Theodore N. Tomaras\footnote{tomaras@physics.uoc.gr}}

\date{\today}


\address{Department of Physics and Institute of Plasma Physics,
 University of Crete, 71003 Heraklion, Greece}

\begin{abstract}

Thin gravitating defects with conical singularities in higher
codimensions and with generalized Israel matching conditions are
known to be inconsistent for generic energy-momentum. A way to
remove this inconsistency is proposed and is realized for an axially
symmetric gravitating codimension-two defect in six dimensional
Einstein gravity. By varying with respect to the brane embedding
fields, alternative matching conditions are derived, which are
generalizations of the Nambu-Goto equations of motion of the defect,
consistent with bulk gravity. For a maximally symmetric defect the
standard picture is recovered. The four-dimensional perfect fluid
cosmology coincides with conventional FRW in the case of radiation,
but for dust it has $\rho^{4/3}$ instead of $\rho$. A
four-dimensional black hole solution is presented having the
Schwarzschild form with a short-distance $r^{-2}$ correction.

\end{abstract}

\maketitle

The study of higher codimension gravitating defects is important for
many reasons. They arise naturally in higher dimensional gravity
theories such as 10-dimensional supergravity or M-theory, they are
relevant to the discussion of braneworld models of particle physics
and cosmology, while the well-known cosmic strings of
four-dimensional GUTs are examples of such codimension-two objects.
In particular, the codimension-two braneworlds in six-dimensional
bulk have also attracted considerable interest recently in relation
to the cosmological constant problem \cite{burgess}.
\par
Here, we shall focus on the study of codimension-two gravitating
defects with zero thickness. They are interesting on their own right
as descriptions of D-braneworld models, and furthermore, they should
be good approximations of model-dependent thick defects away from
their cores \cite{boisseau}. However, in the standard treatment, one
varies the action with respect to the metric; this leads to bulk
Einstein equations with the localized brane energy-momentum tensor
as source proportional to $\delta^{(2)}$ on the right-hand side,
which are known to be inconsistent for generic energy-momentum on
the defect \cite{geroch}. In this paper instead, to obtain the
embedding conditions of the defect, we shall vary the action with
respect to the brane position in a way that takes into account the
gravitational back-reaction. Our approach is reminiscent of the
``Dirac style'' variation performed in \cite{davidson} in the study
of codimension-one defects.
\par
We consider the total brane-bulk action \bea
&&\!\!\!\!\!\!S=\frac{1}{2\kappa_{6}^{2}}\!\int_{\!M}
\!\!d^6\!x\sqrt{|g|}\,
(\mathcal{R}-2\Lambda_{6}\!)+\!\int_{\!\Sigma} \!d^4\!x\sqrt{|h|}\,
\Big(\!\frac{r_{c}^{2}}{2\kappa_{6}^{2}}R-\lambda\Big)\nn\\
&&\,\,\,\,\,\,+\!\int_{\!M}\!\!d^6\!x\,\mathcal{L}_{mat}+\!\int_{\!\Sigma}\!d^{4}\!x\,L_{mat},
\label{6action} \eea where $g_{\mu\nu}$ is the bulk metric tensor,
$h_{\mu\nu}$ is the induced metric on the brane ($\mu,\nu,...$ are
six-dimensional coordinate indices), $\mathcal{R}$ and $R$ are the
bulk and brane scalars, $\lambda$ is the brane tension and $r_{c}$
is the induced-gravity crossover length scale. $\mathcal{L}_{mat}$,
$L_{mat}$ are the matter Lagrangians of the bulk and of the brane.
The bulk equations are \be \mathcal{G}_{\mu\nu}=\kappa_{6}^2
\mathcal{T}_{\mu\nu}-\Lambda_{6} g_{\mu\nu},\label{6eqs} \ee where
$\mathcal{G}_{\mu\nu}$ is the bulk Einstein tensor and
$\mathcal{T}_{\mu\nu}$ is a regular bulk energy-momentum tensor. In
addition, we consider $n_{\alpha}^{\,\,\,\mu}$ ($\alpha=1,2$)
arbitrary unit vectors normal to the brane. The induced metric
$h_{\mu\nu}$ is given by
$h_{\mu\nu}=g_{\mu\nu}-n_{\alpha\mu}n^{\!\alpha}_{\,\,\,\nu}$, or
equivalently, in the basis $(\partial_{i},n_{\alpha})$ by
$h_{ij}=g_{\mu\nu}\,x^{\mu}_{\,,i}\,x^{\nu}_{\,,j}$ ($i,j,...$ are
coordinate indices on the brane).
\par
The ordinary way to consider the interaction of the brane with the
bulk is through the variation $\delta g_{\mu\nu}$ at the position of
the brane, in which case  one should add on the right-hand side of
equation (\ref{6eqs}) the term $\kappa_{6}^2\,
\tilde{T}_{\mu\nu}\,\delta^{(2)}$, where
$\tilde{T}_{\mu\nu}=T_{\mu\nu}\!-\!\lambda\,
h_{\mu\nu}\!-\!(r_{c}^{2}/\kappa_{6}^2)G_{\mu\nu}$. $T_{\mu\nu}$ is
the brane energy-momentum tensor, $G_{\mu\nu}$ the brane Einstein
tensor and $\delta^{(2)}$ the two-dimensional delta function. The
presence of $\delta^{(2)}$ in the resulting equation leads to
matching conditions from the distributional terms with the known
inconsistencies.
\par
Here, we propose that the interaction of the brane with the bulk is
obtained by varying the action with respect to $\delta x^{\mu}$, the
embedding fields of the brane position. In this case, $\delta
g_{\mu\nu}\!=\!-\pounds_{\!\delta x}g_{\mu\nu}$. $\delta h_{ij}$ and
$\delta n_{\alpha\mu}$ could also be expressed in terms of $\delta
x^{\mu}$. However, we find it more convenient to vary $h_{ij}$ and
$n_{\alpha\mu}$ independently by including corresponding Lagrange
multipliers. So, the relation of $h_{ij}$ to $g_{\mu\nu}$ and the
orthonormality of $n_{\alpha}^{\,\,\,\mu}$ require the addition to
$S$ of the following constraint action \bea &&
\!\!\!\!\!\!\!\!\!\!\!\!\!\!\!
S_{c}=\int_{\!\Sigma}\!d^{4}\!x\,\sqrt{|h|}\,\,[\lambda^{ij}(h_{ij}
-g_{\mu\nu}\,x^{\mu}_{\,,i}\,x^{\nu}_{\,,j})\nn\\
&&\,\,\,\,\,\,\,\,\,\,\,\,\,\,\,\,\,+\lambda^{\alpha
i}n_{\alpha\mu}x^{\mu}_{\,,i}+\lambda^{\alpha\beta}
(g_{\mu\nu}n_{\alpha}^{\,\,\,\mu}n_{\beta}^{\,\,\,\nu}-\delta_{\alpha\beta})],
\eea where $\lambda^{ij},\lambda^{\alpha i},\lambda^{\alpha\beta}$
are Lagrange multipliers. Variation of $S\!+\!S_{c}$ with respect to
$n_{\alpha\mu},h_{ij},g_{\mu\nu}$ gives \bea &&
\!\!\!\!\!\!\!\!\delta
(S\!+\!S_{c})|_{\!b\!r\!a\!n\!e}=\int_{\!\Sigma}d^{4}\!x\sqrt{|h|}(\lambda^{\alpha
i}x^{\mu}_{\,,i}+2\lambda^{\alpha\beta}n_{\beta}^{\,\,\,\mu})\delta
n_{\alpha\mu}\nn\\
&&
\!\!\!\!\!\!\!\!\!\!+\!\!\int_{\!\Sigma}\!d^{4}\!x\sqrt{|h|}\Big[\!\lambda^{ij}
\!+\!\frac{1}{2}(T^{ij}\!-\!\lambda
h^{ij}\!)\!-\!\frac{r_{c}^{2}}{2\kappa_{6}^{2}}G^{ij}\!\Big]
\!\delta h_{ij}\nn\eea \bea
&&\!\!\!\!\!\!\!\!\!\!+\!\!\int_{\!\Sigma}\!d^{4}x
\sqrt{|h|}(\lambda^{\alpha\beta}n_{\alpha}^{\,\,\,\mu}n_{\beta}^{\,\,\,\nu}
\!-\!\lambda^{ij}x^{\mu}_{\,,i}x^{\nu}_{\,,j})
\delta g_{\mu\nu}\nn\\
&&\!\!\!\!\!\!\!\!\!\!+\frac{1}{2\kappa_{\!6}^{2}}\!\int_{\!M}\!\!\!d^{6}\!x\sqrt{\!|g|}
\!\Big[\!(\mathcal{G}_{\!\mu\nu}\!-\!\kappa_{6}^{2}\mathcal{T}_{\!\mu\nu}\!)\delta
g^{\mu\nu}\!\!+\!2g^{\mu[\kappa}\!g^{\lambda]\nu}\!(\!\delta
g_{\nu\kappa}\!)_{\!;\lambda\mu}\!\Big]\!\Big|_{\!b\!r\!a\!n\!e}\!.\label{variation}\eea
When $r_{c}\!\!\neq\!\! 0$, one should add in (\ref{6action}) the
integral of the extrinsic curvature $k$ of $\partial\Sigma$; this,
in general, does not affect the dynamics of $\Sigma$ \cite{guven}.
\par
For simplicity, we restrict ourselves to the axially symmetric bulk
ansatz \vspace{-0.2cm} \be
ds_{6}^2=d\varrho^2+L^2(x,\varrho)d\varphi^2+h_{ij}(x,\varrho)dx^{i}dx^{j}.\label{6metric}
\vspace{-0.1cm}\ee $h_{ij}(x,0)$ is the braneworld metric, assumed
to be regular everywhere with the possible exception of isolated
singular points \cite{moss}. $\varphi$ has the standard periodicity
$2\pi$ and we make the usual assumption for conical singularities
$L(x,\varrho)=\beta(x)\varrho+\mathcal{O}(\varrho^2)$ for
$\varrho\approx 0$, $L'(x,0)=1$, where a prime denotes
differentiation with respect to $\varrho$. The non-vanishing
components of $\mathcal{G}_{\mu\nu}$ are \bea &&
\!\!\!\!\!\!\mathcal{G}_{ij}\!=\!G_{ij}\!+\!\frac{L''}{L}h_{ij}\!-\!\frac{1}{2}h_{ij}''
\!+\!\frac{1}{2}h_{ik}'h_{j\ell}'h^{k\ell}\!-\!\frac{1}{4}h^{k\ell}h_{k\ell}'h_{ij}'\nn\\
&&\!\!\!\!\!\!+\frac{1}{2}h^{k\ell}h_{k\ell}''h_{ij}
\!+\!\frac{3}{8} h^{k\ell\prime} h_{k\ell}'
h_{ij}\!+\!\frac{1}{8}(h^{kl}
h_{kl}')^{2} h_{ij}\nn\\
&&\!\!\!\!\!\!+\frac{1}{L^{2}} (\mathcal{R}_{\varphi i \varphi
j}\!-\!\mathcal{R}_{\varphi k \varphi
\ell}h^{k\ell}h_{ij})\label{uno}\\
&&\!\!\!\!\!\!\mathcal{G}_{\varrho\varrho}\!=\!-\frac{1}{2}R\!+\!\frac{1}{8}h^{k\ell\prime}h_{k\ell}'
\!+\!\frac{1}{8}(h^{k\ell}h_{k\ell}')^{2}\!-\!\frac{1}{L^{2}}h^{k\ell}\mathcal{R}_{\varphi
k \varphi
\ell}\label{dos}\\
&&\!\!\!\!\!\!\mathcal{G}_{\varphi\varphi}\!=\!-\frac{L^{2}}{2}\Big[\!R\!-\!h^{k\ell}h_{k\ell}''\!-\!\frac{3}{4}
h^{k\ell\prime}h_{k\ell}'\!-\!\frac{1}{4}(h^{k\ell}h_{k\ell}')^{2}\!\Big]
\label{express} \\
&&\!\!\!\!\!\!\mathcal{G}_{\varrho
i}=\frac{1}{L^{2}}\mathcal{R}_{\varphi\varrho\varphi
i}\!+\!h^{k\ell}\mathcal{R}_{\varrho ki\ell},\label{gamboa}\eea
where $\mathcal{R}_{\varphi i \varphi
j}\!=\!L(L_{,k}\Gamma^{k}_{\,\,ij}\!-\!L_{,ij}\!-\!\frac{1}{2}L'h_{ij}')$,
$\mathcal{R}_{\varphi\varrho\varphi
i}=L(\frac{1}{2}L_{,k}h^{k\ell}h_{\ell i}'-L'_{,i})$, while
$\mathcal{R}_{\varrho ijk}$ does not contain $L$. Integrate over the
$(\varrho, \varphi)$ transverse disc of radius $\epsilon$ the
six-dimensional terms of (\ref{variation}) in the limit
$\epsilon\rightarrow 0$. For smooth $\delta g_{\mu\nu}$, the only
term in the above components of $\mathcal{G}_{\mu\nu}$ which
contributes is $(L''/L)h_{ij}$, having the most singular
$\delta(\varrho)/\varrho$ structure. The result is \bea&&
\!\!\!\!\!\!\delta
(S\!+\!S_{c})|_{\!b\!r\!a\!n\!e}=\int_{\!\Sigma}d^{4}\!x\sqrt{|h|}(\lambda^{\alpha
i}x^{\mu}_{\,,i}+2\lambda^{\alpha\beta}n_{\beta}^{\,\,\,\mu})\delta
n_{\alpha\mu}\nn\\
&&
\!\!\!\!\!\!+\!\!\int_{\!\Sigma}\!\!d^{4}\!x\sqrt{\!|h|}\!\Big[\!\lambda^{ij}
\!\!+\!\frac{1}{2}(T^{ij}\!\!-\!\lambda h^{ij}\!)
\!+\!\frac{\pi}{\kappa_{6}^{2}}(1\!-\!\beta)h^{ij}\!\!-\!\frac{r_{c}^{2}}{2\kappa_{6}^{2}}G^{ij}\!\Big]\!\delta
h_{ij}\nn\\
&&\!\!\!\!\!\!+\!\!\int_{\!\Sigma}\!d^{4}x
\sqrt{|h|}(\lambda^{\alpha\beta}n_{\alpha}^{\,\,\,\mu}n_{\beta}^{\,\,\,\nu}
\!-\!\lambda^{ij}x^{\mu}_{\,,i}x^{\nu}_{\,,j})\delta
g_{\mu\nu}.\label{final variation} \eea At this point, had we
considered $\delta g_{\mu\nu},\delta h_{ij},\delta n_{\alpha\mu}$
independent, all Lagrange multipliers would have been zero, leading
to the matching conditions
$\kappa_{6}^{2}T_{\mu\nu}=[\kappa_{6}^{2}\lambda-2\pi
(1\!-\!\beta)]h_{\mu\nu}+r_{c}^{2}G_{\mu\nu}$, which are
incompatible with any $T_{\mu\nu}$ other than a brane tension
\cite{rc}. Instead, as explained above, our variation is with
respect to $x^{\mu}$. So, $\delta
(S\!+\!S_{c})|_{\!b\!r\!a\!n\!e}=0$ gives \bea
&&\!\!\!\!\!\!\!\!\!\!\!\!\!\!\!\!\lambda^{\alpha
i}x^{\mu}_{\,,i}+2\lambda^{\alpha\beta}n_{\beta}^{\,\,\,\mu}=0\label{ena}\\
&&\!\!\!\!\!\!\!\!\!\!\!\!\!\!\!\!
\lambda^{ij}=\frac{1}{2\kappa_{6}^{2}}\{[\kappa_{6}^{2}\lambda\!-\!2\pi(1\!-\!\beta)]h^{ij}\!-\!\kappa_{6}^{2}T^{ij}\}
+\frac{r_{c}^{2}}{2\kappa_{6}^{2}}G^{ij}\label{dio}\\
&&\!\!\!\!\!\!\!\!\!\!\!\!\!\!\!\!\int_{\!\Sigma}\!d^{4}x
\sqrt{|h|}(\lambda^{\alpha\beta}n_{\alpha}^{\,\,\,\mu}n_{\beta}^{\,\,\,\nu}
\!-\!\lambda^{ij}x^{\mu}_{\,,i}x^{\nu}_{\,,j})\delta
g_{\mu\nu}=0\label{trii}, \label{equations} \eea where $\delta
g_{\mu\nu}\!=\!-\pounds_{\!\delta x}g_{\mu\nu}
\!=\!-(g_{\mu\nu,\lambda}\delta x^{\lambda}+g_{\mu\lambda}\delta x^{
\lambda}_{\,,\nu}+g_{\nu\lambda}\delta x^{\lambda}_{\,,\mu})$. Since
$x^{\mu}_{\,,i},n_{\alpha}^{\,\,\,\mu}$ are independent, equation
(\ref{ena}) implies $\lambda^{\alpha
i}\!=\!\lambda^{\alpha\beta}\!=\!0$. Then, equation (\ref{trii})
with $\lambda_{ij}$ given by equation (\ref{dio}) takes the form
\bea
\!\!\int_{\!\Sigma}\!d^{4}x\sqrt{|h|}\lambda^{ij}(g_{\mu\nu,\lambda}x^{\mu}_{\,,i}x^{\nu}_{\,,j}\delta
x^{\lambda}\!+\!2g_{\mu\nu}x^{\mu}_{\,,i}x^{\lambda}_{\,,j}\delta
x^{\nu}_{\,,\lambda})\!=\!0, \label{tria1} \eea which, after an
integration by parts and imposing $\delta x^{\mu}|_{\partial
\Sigma}=0$, becomes \bea
\!\!\int_{\!\Sigma}\!\!d^{4}\!x\sqrt{|h|}\,g_{\mu\sigma}[\lambda^{ij}_{\,\,\,|j}x^{\mu}_{\,,i}
\!+\!\lambda^{ij}(x^{\mu}_{\,\,;ij}\!+\!\Gamma^{\mu}_{\,\,\,\nu\lambda}x^{\nu}_{\,,i}x^{\lambda}_{\,,j})]\delta
x^{\sigma}\!\!=\!0. \label{step} \eea In (\ref{step}), $|$ and $;$
denote covariant differentiations corresponding to $h_{ij}$ and
$g_{\mu\nu}$ respectively. Due to the arbitrariness of $\delta
x^{\mu}$ and since the extrinsic curvatures of the brane $K_{\alpha
ij}\!=\!n_{\alpha i;j}$ satisfies
$-K^{\alpha}_{\,\,\,\,ij}n_{\alpha}^{\,\,\,\mu}=
x^{\mu}_{\,\,;ij}+\Gamma^{\mu}_{\,\,\,\nu\lambda}x^{\nu}_{\,,i}x^{\lambda}_{\,,j}$,
the last equation is equivalent to
$\lambda^{ij}_{\,\,\,|j}x^{\mu}_{\,,i}
-\lambda^{ij}K^{\alpha}_{\,\,\,\,ij}n_{\alpha}^{\,\,\,\mu}=0$, from
which
$\lambda^{ij}_{\,\,\,|j}=\lambda^{ij}K^{\alpha}_{\,\,\,\,ij}=0$.
Finally, using (\ref{dio}) we obtain \bea
&&\!\!\!\!\!\!\!\!\!\!\!\!\!\!\!\!T^{ij}_{\,\,\,\,|j}
=\frac{2\pi}{\kappa_{6}^{2}}h^{ij}\beta_{,j} \label{1}\\
&&
\!\!\!\!\!\!\!\!\!\!\!\!\!\!\!\!\!\Big{\{}\!\kappa_{6}^{2}T^{ij}-[\kappa_{6}^{2}\lambda\!-\!2\pi
(1\!-\!\beta)]h^{ij}-r_{c}^{2}G^{ij}\!\Big{\}}K^{\alpha}_{\,\,\,\,ij}\!=\!0.
 \label{2} \eea
Equation (\ref{2}) is the generalization of the Nambu-Goto equation
of motion, when the self-gravitating brane interacts with bulk
gravity. Indeed, in the special case of a probe brane of tension
$\lambda$ (no back-reaction, $T_{ij}=0$, $r_{c}=0$), equation
(\ref{2}) becomes $h^{ij}K^{\alpha}_{\,\,\,\,ij}=0$, or equivalently
$\Box_{h}x^{\mu}\!+\!\Gamma^{\mu}_{\,\,\,\nu\lambda}h^{\nu\lambda}=0$,
which is the Nambu-Goto equation of motion \cite{box}.
\par
We will now examine all the bulk field equations (\ref{6eqs}) at the
position of the brane and check their compatibility with the
matching conditions (\ref{1}), (\ref{2}). Focusing on the
$\mathcal{O}(1/\varrho)$ terms in the $\varrho i$ components of
equations (\ref{6eqs}) (which cannot be canceled by any regular
$\mathcal{T}_{\mu\nu}$ in (\ref{6eqs})) we obtain $\beta_{,i}=0$,
and equation (\ref{1}) gives the exact conservation on the brane
\vspace{-0.2cm}\be T^{ij}_{\,\,\,\,\,|j}=0. \label{exact}
\vspace{-0.2cm}\ee Similarly, from the $\mathcal{O}(1/\varrho)$ part
of the $\varrho\varrho$ component of (\ref{6eqs}) we obtain
$h^{ij}h_{ij}'=0$, valid at the position of the brane. The only
nontrivial remaining components of (\ref{6eqs}) with a
$\mathcal{O}(1/\varrho)$ part are the $ij$ ones, which give on the
brane the equation
$\widehat{L''}\,h_{ij}+\frac{\beta}{2}(h^{k\ell}h_{k\ell}'h_{ij}-h_{ij}')=0$,
with a hat denoting the regular part of the corresponding quantity.
This equation is equivalent to
$\widehat{L''}\,h_{ij}=\frac{\beta}{2}h_{ij}'$, which means that on
the brane $h_{ij}'=0$. Thus, \be K^{\alpha}_{\,\,\,\,ij}=0,
\label{tri} \ee which trivially satisfies (\ref{2}), and from which
one obtains for the brane the geodesic equation
$x^{\mu}_{\,\,;ij}\!+\!\Gamma^{\mu}_{\,\,\,\nu\lambda}x^{\nu}_{\,,i}x^{\lambda}_{\,,j}=0$.
\par
A few comments are in order at this point: First, it may be natural
to expect that in the ``strong probe limit'' defined as the limit
$\tilde{T}_{\mu\nu}\rightarrow 0$, the brane obeys the geodesic
equation. This is trivially the case with the axially symmetric
codimension-two defects, as described above by the geodesic equation
(\ref{tri}). Incidentally, the codimension-one Israel matching
conditions with $Z_{2}$-symmetry in $D$-dimensions
$2K_{ij}\!=\!-\kappa_{D}^{2}[\tilde{T}_{ij}-\tilde{T}h_{ij}/(D\!-\!2)]$
satisfy this expectation. On the contrary, the codimension-one
matching condition arising by variation with respect to $x^{\mu}$,
$\{2K_{ij}+\kappa_{D}^{2}[\tilde{T}_{ij}-\tilde{T}h_{ij}/(D\!-\!2)]\}K^{ij}\!=\!0$
does not. Second, the six-dimensional curvature invariants at the
position of the brane contain singular $\delta(\varrho)/\varrho$
terms dealt above, and, in general, also terms containing powers of
$1/\varrho$ multiplied by $h_{ij}^{\,\prime}$, which vanish because
$h_{ij}^{\,\prime}=0$. This means that the bulk geometry is regular
at the position of the brane. Third, in the general non-axially
symmetric case, the bulk equations will not necessarily imply the
geodesic motion for the defect and the curvature will be divergent
at its position \cite{israel}. Furthermore, the corresponding
matching conditions are not expected to be trivially satisfied.
\par
Finally, we focus on the regular part of (\ref{6eqs}) on the brane.
The $\varrho\varrho$ component gives \be
R=2\Lambda_{6}-2\kappa_{6}^{2}\mathcal{T}_{\varrho\varrho}.
\label{radio} \ee Similarly, from the $\varrho i$ components one
obtains $h^{k\ell}\mathcal{R}_{\varrho
ki\ell}=\kappa_{6}^{2}\mathcal{T}_{\varrho i}$, or equivalently
$h^{k\ell}(K_{\varrho ki\,!\,\ell}-K_{\varrho
k\ell\,!\,i})=\kappa_{6}^{2}\mathcal{T}_{\varrho i}$. At this point,
we have used the geometric identity
$\mathcal{R}^{\alpha}_{\,\,\,\,ijk}=K^{\alpha}_{\,\,\,\,ij\,!\,k}-K^{\alpha}_{\,\,\,\,ik\,!\,j}$,
where the covariant derivative $!$ is defined with respect to the
connection $\varpi_{\beta\alpha
i}=g(\nabla_{i}n_{\alpha},n_{\beta})$ as
$\Phi^{\alpha}_{\beta\,!\,i}=\Phi^{\alpha}_{\beta|i}+\varpi^{\alpha}_{\,\,\,\gamma
i}\Phi^{\gamma}_{\beta}- \varpi^{\gamma}_{\,\,\,\beta
i}\Phi^{\alpha}_{\gamma}$ for fields $\Phi^{\alpha}_{\beta}$
transforming as tensors under normal frame rotations
$\Phi^{\alpha}_{\beta}\!\rightarrow\!
O_{\beta}^{\,\,\,\delta}(O^{-1})_{\gamma}^{\,\,\,\alpha}\Phi_{\delta}^{\gamma}$
(the derivative $|$ in the previous expression is meant on
tangential indices $i,j,...$ that $\Phi^{\alpha}_{\beta}$ may
possess). Therefore, due to (\ref{tri}), the previous $\varrho i$
equation is identically satisfied whenever $\mathcal{T}_{\varrho
i}=0$ on the brane. Concerning the regular part $\varphi\varphi$ of
(\ref{6eqs}), and using (\ref{tri}), (\ref{radio}) one obtains \be
h^{ij}\widehat{h_{ij}''}=2\kappa_{6}^{2}(\mathcal{T}^{\varphi}_{\varphi}-\mathcal{T}^{\varrho}_{\varrho}).
\label{nama} \ee Finally, using (\ref{nama}), the $ij$ part becomes
 \be
\widehat{h_{ij}''}=2G_{ij}+2\Lambda_{6}h_{ij}-2\kappa_{6}^{2}[\mathcal{T}_{ij}\!-\!
(\mathcal{T}^{\varphi}_{\varphi}\!-\!\mathcal{T}^{\varrho}_{\varrho})h_{ij}].
\label{spa} \ee Combining (\ref{radio}), (\ref{nama}) and
(\ref{spa}) one gets on the brane
$2\Lambda_{6}=\kappa_{6}^{2}(\mathcal{T}^{i}_{i}\!+\!\mathcal{T}^{\varrho}_{\varrho}
\!-\!3\mathcal{T}^{\varphi}_{\varphi})$ and  \be
R=\kappa_{6}^{2}(\mathcal{T}^{i}_{i}\!-\!\mathcal{T}^{\varrho}_{\varrho}\!-\!3\mathcal{T}^{\varphi}_{\varphi}).
\label{ygt} \ee \par In summary, the brane-bulk system is described
by the regular piece of equations (\ref{6eqs}), under the boundary
conditions (\ref{exact}), (\ref{tri}) and (\ref{ygt}). In the
absence of bulk matter $\mathcal{T}_{\mu\nu}=0$, one gets
$\Lambda_{6}=0$ \cite{LinetB}, and (\ref{ygt}) reduces to $R=0$.
This Ricci scalar flat condition is enough in special cases of high
symmetry to determine the brane geometry.
\par
We now proceed with the discussion of a few cases with physical
interest: The maximally symmetric solution of equation $R=0$ is the
Minkowski brane. The bulk solution which is consistent with this and
(\ref{tri}) is the locally flat geometry with a deficit angle
\cite{vilenkin}
$ds_{6}^{2}=d\varrho^{2}+\beta^{2}\varrho^{2}d\varphi^{2}-dt^{2}+d\vec{x}^{2}$,
where $\beta$ is an integration constant \cite{kofinas}. Here, we do
not have equation $\mathcal{G}_{\mu\nu}=-\kappa_{6}^{2}\lambda
h_{\mu\nu}\delta^{(2)}$ to obtain the standard relation
$\kappa_{6}^{2}\lambda=2\pi(1\!-\!\beta)$ \cite{levi}. The latter we
would like to arise as a direct calculation of an appropriately
defined energy of the gravitational field. Indeed, one such
definition that has successfully passed various other tests is given
by the teleparallel representation of Einstein gravity described in
\cite{maluf}, and gives the energy per unit length of the defect of
the gravitational field of our bulk solution inside a cylinder of
arbitrary radius around the defect exactly equal to
$(1\!-\!\beta)/4$.
\par
Next, we consider the bulk cosmological metric of the form
$ds_{6}^2\!=\!d\varrho^{2}\!+\!L^{2}(t,\varrho\!)d\varphi^{2}\!-\!n^{2}(t,\varrho\!)dt^{2}\!+\!a^{2}(t,\varrho\!)
\gamma_{\hat{i}\hat{j}}(x)dx^{\hat{i}}\!dx^{\hat{j}}$, where
$\gamma_{\hat{i}\hat{j}}$ is a maximally symmetric 3-dimensional
metric characterized by its spatial curvature $k=-1,0,1$. For this
metric, equation $R=0$ is $n^{-1}\dot{H}+2H^{2}+ka^{-2}=0$, with
$H\!=\!\dot{a}/na$ being the Hubble parameter of the brane and a dot
denotes differentiation with respect to $t$. The lapse function
$n(t,0)$ can be left undetermined, since it corresponds to the
temporal gauge choice. The matter on the brane is taken to be a
perfect fluid with energy density $\rho$ and pressure $p\!=\!w\rho$,
in which case, the conservation equation (\ref{exact}) gives
$\rho=Ma^{-3(1+w)}$, with $M$ an integration constant. The equation
for $H$ is integrated to $H^{2}=c_{1}a^{-4}\!-\!ka^{-2}$, with
$c_{1}$ integration constant, or writing in terms of $\rho$,
$H^{2}=c\rho^{4/3(1+w)}-ka^{-2}$, with $c\!=\!c_{1}M^{-4/3(1+w)}$.
Since in the matter era the Hubble parameter falls faster than in
the standard scenario, the age of the universe is predicted to be
the 3/4 of the one given in the standard cosmology. Moreover, since
the radiation era is left unchanged, the hot big bang predictions
remain intact. Note that a non-vanishing $\mathcal{T}_{\mu\nu}$
would give a contribution to the effective cosmological constant,
that could drive an inflationary or a late-time acceleration era.
Indeed, curiously enough, the fitting \cite{leandros} to the
supernovae data of the previous Hubble evolution gave $\chi^{2}=175$
(about $1\sigma$ better than in $\Lambda$CDM) and $\Omega_{m}=0.17$.
However, a full six-dimensional treatment would be needed to study
perturbative aspects of this cosmology, such as CMB or structure
formation.
\par
Finally, we consider a static spherically symmetric brane in the six
dimensional bulk of the form
$ds_{6}^2\!=\!d\varrho^{2}\!+\!L^{2}(r,\varrho\!)d\varphi^{2}
\!-\!A(r,\varrho\!)dt^{2}\!+\!B(r,\varrho\!)dr^{2}+C(r,\varrho)d\Omega_{2}^{2}$,
where we can use the gauge choice on the brane $C(r,0)=r^{2}$ to
bring the four-dimensional metric in the standard form. There are
two functions to be determined on the brane: $A(r,0),B(r,0)$.
However, the $R=0$ equation is not enough to determine both. To
close the system, consider as an example the ansatz on the brane
$B(r,0)=A(r,0)^{-1}$. Then, $A(r)=A(r,0)$ satisfies
$r^{2}A''+4rA'+2A-2=0$, with the general solution
$A(r)=1+c_{1}r^{-1}+c_{2}r^{-2}$ ($c_{1},c_{2}$ integration
constants). This is the standard Schwarzschild metric with a short
distance $r^{-2}$ correction term.
\par
In analogy with the present treatment of axial symmetry, the
addition of a Gauss-Bonnet term in the bulk \cite{ruth} would lead
to the resulting matching condition \bea &&\!\!\!\!\!\!
\Big\{\!K^{\beta\ell}_{\,\,\,\,\,\,\ell} K_{\beta ij}\!-\!
K^{\beta\ell}_{\,\,\,\,\,\,i}K_{\beta j\ell} \!+\!\frac{1}{2}(
K^{\beta k\ell}K_{\beta k\ell}\!-\!
K^{\beta k}_{\,\,\,\,\,\,k}K_{\beta\,\,\,\ell}^{\,\,\,\,\ell})h_{ij}\nn\\
&&\!\!\!\!\!\!\!+\!\Big(\!\!\beta^{-\!1}\!\!-\!\!1\!\!+\!\frac{r_{c}^2}{8\pi\alpha\beta}\!\Big)\!G_{\!ij}\!+\!
\frac{\kappa_{6}^{2}\lambda\!\!-\!\!2\pi(\!1\!\!-\!\!\beta)}{8\pi\alpha\beta}h_{\!ij}\!\!-\!\!
\frac{\kappa_{6}^{2}}{8\pi\alpha\beta}T_{\!ij}\!\Big\}\!K^{\alpha
ij}\!\!=\!0,\nn \eea obtained under the assumptions of discontinuous
$L'$ and $h_{ij}'$ ($h_{ij}'(x,0)\!=\!0\!\neq\!h_{ij}'(x,0^{+})$).
In this case, $\beta$  in general will not be constant, and equation
(\ref{tri}) will not be true any longer, which means that the
matching condition is not trivially satisfied. By counting, however,
the degrees of freedom and the equations we find an agreement, thus,
we still expect that the whole bulk system will be compatible with
this new matching condition leading to a final complicated
braneworld dynamics.

\par
To summarize, we studied a codimension-two zero-thickness
brane in the context of six-dimensional Einstein gravity, possibly extended
by the addition of an induced gravity term on the brane. Our discussion
focused on the axially symmetric bulk geometry.

It is well known that the analogue of the Israel matching conditions
of such braneworlds (obtained by varying the action with respect to
the brane metric) are inconsistent for any matter content on the
brane, with the exception of just brane tension. Here, alternative
matching conditions for the brane are proposed, obtained instead, by
varying the action with respect to the brane position. They are
generalizations of the Nambu-Goto equations of motion of a probe
brane, to take into account the gravitational back-reaction. They
are compatible with the full set of field equations at the position
of the brane, and according to these, the brane motion in the bulk
is geodesic.

The geometry on the brane itself is constrained by a vanishing
induced Ricci scalar, plus the standard energy-momentum conservation
equation on the brane. In simple cases of high symmetry these
conditions are enough to determine the geometry. We analyzed the
following such characteristic cases of physical interest. For a
maximally symmetric defect the standard picture of a locally flat
geometry with the right deficit angle is recovered. For an isotropic
perfect fluid four-dimensional cosmology, the Friedmann equation
derived coincides with the conventional cosmology in the radiation
era, while in the matter-dominated era it behaves like $\rho^{4/3}$
instead of $\rho$. For a four-dimensional spherically symmetric
braneworld with the ansatz $g_{tt}=g_{rr}^{-1}$, the geometry on the
brane is the Schwarzschild one with a short distance $r^{-2}$
correction term.

Extending the present analysis to a generic - less symmetric - bulk
is expected to lead to more complicated dynamics, able to address
also the issue of a fluctuating defect. Generalization to other
gravitational theories and study of even higher-codimension
defects of string theory would be extensions of the present work.

\vspace{-0.3cm}

\[ \]

 {\bf Acknowlegements}. We wish to thank C.
Charmousis, A. Davidson, V. Frolov, R. Maartens, G. Panotopoulos, L.
Perivolaropoulos and A. Petkou for useful discussions.


\begin{references}
\bibitem{rc} This is obvious for $r_{c}\!=\!0$. For $r_{c}\!\neq
\!0$ one has to use also equations (\ref{6eqs}) at the brane
position to obtain incompatibility (see equation (\ref{radio})).
\bibitem{box} Note that the vector
$\Box_{h}x^{\mu}\!+\!\Gamma^{\mu}_{\,\,\,\nu\lambda}h^{\nu\lambda}$
is normal to the brane, so the last expression of the Nambu-Goto
equation with one free index $\mu$ is equivalent to the previous one
with one free index $\alpha$.
\end{references}
\end{document}